# Forensic Analysis of Residual Information in Adobe PDF Files

Hyunji Chung
Center for Information Security Technologies
Korea University
Seoul, Korea
foryou7187@korea.ac.kr

Jungheum Park
Center for Information Security Technologies
Korea University
Seoul, Korea
junghmi@korea.ac.kr

Sangjin Lee
Center for Information Security Technologies
Korea University
Seoul, Korea
sangjin@korea.ac.kr

*Abstract*

*In recent years, as electronic files include personal records and business activities, these files can be used as important evidences in a digital forensic investigation process. In general, the data that can be verified using its own application programs is largely used in the investigation of document files. However, in the case of the PDF file that has been largely used at the present time, certain data, which include the data before some modifications, exist in electronic document files unintentionally. Because such residual information may present the writing process of a file, it can be usefully used in a forensic viewpoint.*

*This paper introduces why the residual information is stored inside the PDF file and explains a way to extract the information. In addition, we demonstrate the attributes of PDF files can be used to hide data.*

1. Introduction

People create their electronic documents using various application programs, such as Microsoft Office, Adobe Acrobat, and so on. Although the use of document files is widespread, few people recognize that 'hidden' data exist in files. The reason for using the word 'hidden' is that identification of these data is not possible using its own application program.

J. Park et al. introduced 'hidden' data in MS PowerPoint files. Even after the contents of an MS PowerPoint file have been deleted or edited, they can still exist inside the file as residual information. This is because file-saving algorithm used in this application[2]. Similar to MS PowerPoint files, Adobe PDF (Portable Document Format) files can also contain 'hidden' data. Therefore, it is necessary to investigate such 'hidden data' in a digital forensic viewpoint.

Adobe PDF files have been largely used for various purposes such as writing personal documents and distributing official documents at enterprise. In particular, for enterprise purposes, some confidential documents were created using applications like Microsoft Word or PowerPoint. And they were distributed after transforming into

Adobe PDF files. In the past, PDF files were just application for distribution. However, the content of PDF files can be edited using various PDF editors in these days.

The following is a hypothetical case related to Adobe PDF files. Company "Y" recruited new business partner for new technologies. As a result, both company "A" and company "B" were on the shortlist. The company "B" bought up a core member of the company "A" in order to modify final proposal and be business partner with company "Y". Finally, the modified file was submitted, and company "B" was selected as a partner. Some three years later, certain suspicious elements were detected in the competitive selecting process, following which investigation of the business was undertaken. The only evidence was the Adobe PDF file created for the final proposal after long time.

In the case, forensic examiners were able to investigate some evidences in Adobe PDF file. This is because the contents have been deleted or edited can exist as residual information inside the file due to the file-update mechanism.

This study consists of seven sections. Section 2 introduces the existing studies on Adobe PDF files. Section 3 describes the internal structure of Adobe PDF files. Section 4 represents the reason that PDF files can include residual information. Section 5 proposes a way to extract the residual information. Section 6 describes a data hiding method using a PDF update mechanism. Finally, Section 7 represents the conclusion of this study.

2. Related studies

In 2008, Didier and Matthew introduced that updated PDF files can include revision history by updating mechanism of Adobe PDF. That is, they explained it is possible to know revision history in a PDF file when the file is modified and saved using 'Save' instead of 'Save As' function[1][5].

Didier was developed tools, such as pdf-parser.py, pdfid.py for analysis of PDF files. The tools help examiners to identify an internal structure of PDF files. However, it is difficult to identify specific ASCII, Unicode text for digital forensic investigation[5].

Also, Matthew was developed a tool, pdfresurrect. The pdfresurrect shows the revision number and time of a PDF file. This tool makes investigators to confirm revision history. However, since it cannot verify what data is modified, there is a limit to utilizing in digital forensic purposes[1].

Furthermore, this study researches about unused area in PDF documents from the viewpoint of digital forensics. This paper introduces a tool for analyzing PDF files. It is developed with the special function of extracting text data(both ASCII and Unicode) at each revision. It is easy to compare modified version with original version. Therefore, the result of this paper is useful for investigating Adobe PDF files. Also, this paper explains that unused area in PDF files can be used for data hiding.

3. Internal structure of PDF files

As shown in Figure 1, a PDF file consists of Header, Body, Cross-reference table (hereinafter called xref), and Trailer[3].

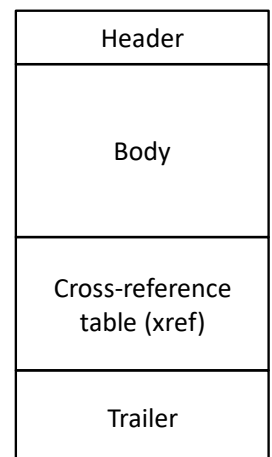

**Figure 1. Basic Structure of PDF files**

The first line of a PDF file is header that has a version number. The trailer is at the end of a PDF file. It has byte offset of the last xref[3].

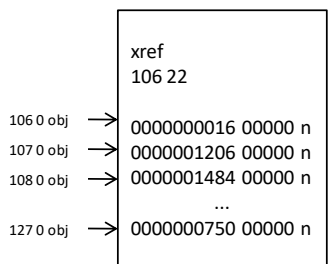

**Figure 2. Example of cross-reference table (xref)**

Xref contains information of all objects in a PDF file. The example of the xref is shown in Figure 2. There can be several xrefs, but Figure 2 shows only one of them. The term of "106 22" means that there exists twenty-two objects from the 106th object. Then, the first ten places in the next line represent the position of the 106th object as a byte offset. The second five places mean the generation number, and the value of '00000' is allocated as it is first generated. For the third place, 'n' or 'f' can be positioned where 'n' is the object, which is being used, and 'f' is the object, which is not used (i.e., free)[3].

Many objects exist in a PDF file. Objects are logically made of a tree. As shown in Figure 3, /Root, /Pages, /Kids, /Page and /Contents represents the type of objects.

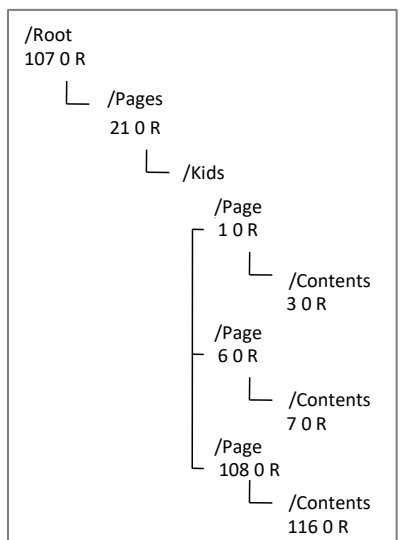

**Figure 3. Internal structure of a PDF file**

Although there are various different types in addition to the type presented in Figure 3, this study refers the main types only. In Figure 3, the root of the internal structure tree is the /Root type. The /Root contains an indirect reference, "107 0 R". In detail, xref in Figure 2 includes byte offset of "107 0 obj". /Pages type exists at byte offset 1206. Continuously, /Pages has indirect reference, "21 0 R". The "21 0 obj" is /Kids type, and it is possible to access the position of "21 0 obj" through proper xref. /Kids consists of one or more indirect /Page references. It is possible to know how many pages PDF file has. This is because a /Page type means each page inside a file. Figure 3 is an example of the PDF file that has three pages. In figure 3, the one of /Page types has indirect reference, "1 0 R". The "1 0 obj" is /Contents type, and the position of it can be accessed using xref. Also, /Contents has indirect reference, "3 0 R". The "3 0 obj" is the body of a page, and the data will be acquired through xref that saves the position as a byte offset.

In addition, the body data can be a compressed(or encoded) format. Thus, it is necessary to decompress(or decode) the data in order to confirm plain text.

## 4. Residual information in PDF files

### 1) Update mechanism of PDF files

Lots of users recently build their electronic documents using specific application programs like Microsoft Office 2007 and store them as PDF files through PDF transformation process. The Adobe Acrobat 8.0 was used in this experiment. Table 1 shows the experiment procedure.

**Table 1. Experiment procedure**

| Step | |
|---|---|
| Step1 | Create three-page contents (texts and images) using Microsoft Office Word 2007 and store it as a PDF file using the function of "Save as Adobe PDF". |
| Step2 | Open the PDF file using Adobe Acrobat. |
| Step3 | Modify the content in the first and second pages of the original PDF file into different texts and images after deleting the existing them. |
| Step4 | Resave it using the function of "Save" after completing the modification. |

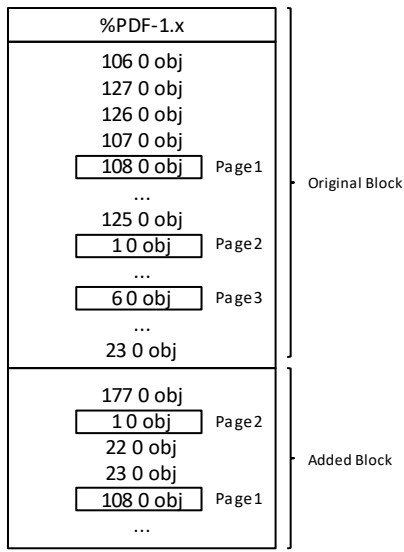

**Figure 4. Structure after completing the experiment**

Figure 4 shows the internal format after finishing the experiment. Repeating Step 3 and Step 4 attaches 'added block' to 'original block'. "108 0 obj" contains contents of the first page. "1 0 obj" has them of the second page and "6 0 obj includes them of the third page.

Both "108 0 obj" and "1 0 obj" exist in 'added block', and "6 0 obj" doesn't exist in 'added block'. This is because the third page is not modified. As shown in Figure 5, it can be seen that the size of the modified file is bigger than the size of original one.

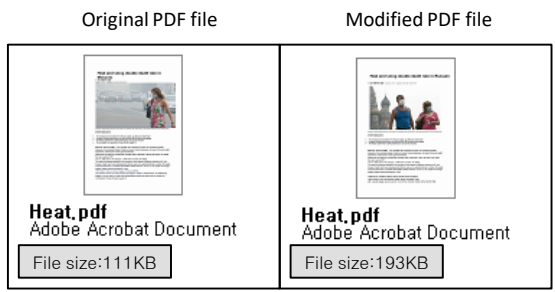

**Figure 5. Comparison of original file size and modified file size**

Residual information is generated when the file is updated. Figure 4 and 5 show this feature. This feature improves the efficiency of saving a PDF file. In other words, it takes less time than a full save of the file. The reason for using 'residual' is that it cannot be identified by PDF application. However, if a user saves the file using the "Save As" function, the application does reconstruct the entire structure.

2) Residual information in PDF files

Figure 6 explains the concept of residual information. In Figure 6, the left side is xref of the modified file, and the right side is the internal structure. "108 0 obj" and "1 0 obj" exist in both original and added blocks. "108 0 obj" in 'added block' is appended after modifying the file. Therefore, "108 0 obj" in 'original block' exist in the file as the residual information.

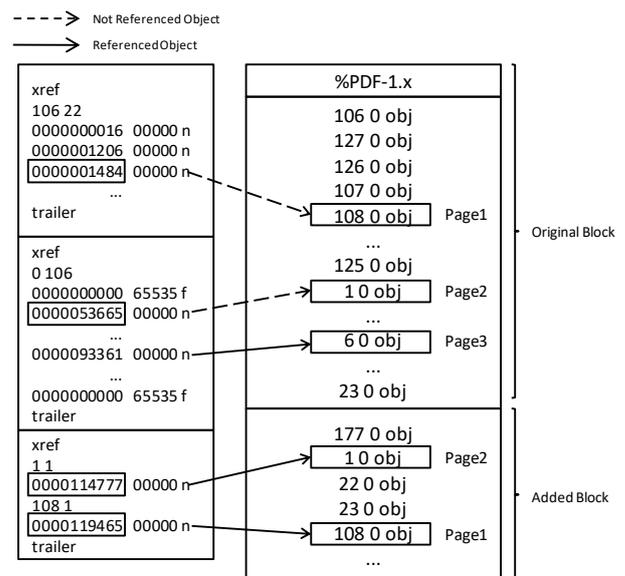

**Figure 6. State of reference in modified file**

As the first and second pages are modified, the objects related to the modification are included in both 'original block' and 'added block'. The first page after the modification does not use the $108^{th}$ object in the 'original block' but use the $108^{th}$ object in the 'added block'. The $108^{th}$ and the $1^{st}$ objects in the 'original block' are not used by the PDF file viewer.

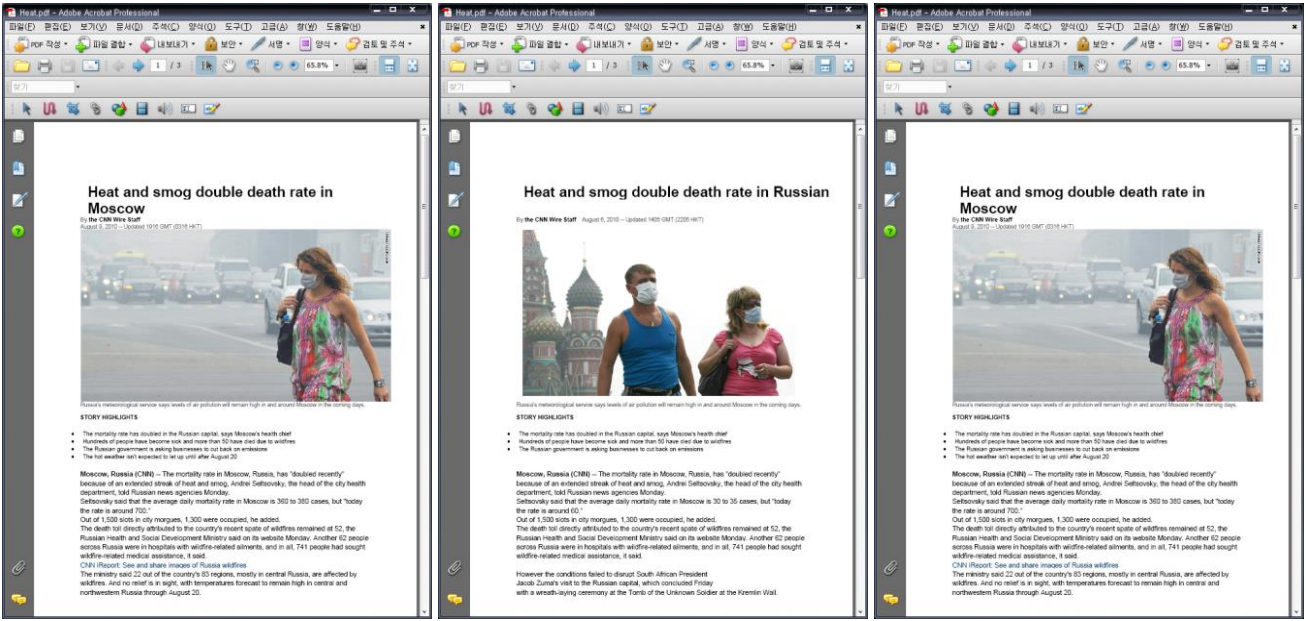

**Figure 7. Size and contents of original file, modified file, and recovered file**

## 5. Extraction of residual information

1) Recovering the file to original version

In figure 6, the first page of the PDF file after the modification uses the 108$^{th}$ object (byte offset 119465) in the 'added block'. Also, the second page uses the 1$^{th}$ object (byte offset 114777) in the 'added block'.

In Figure 8, to access the contents of original file, there is one methodology that it changes byte offset 114777 into byte offset 53665. In addition, it replaces byte offset 119465 with byte offset 1484. Then, the file is recovered to original version. Also, the contents of the original file can be viewed using a PDF file viewer.

Figure 7 shows the contents identified by a PDF file viewer. It is verified that the size of the recovered file is the same as the modified file and shows the same contents as the original file.

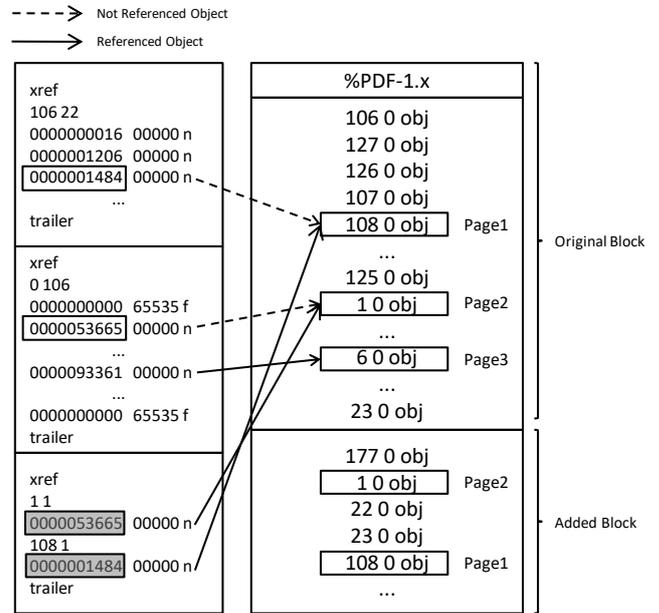

**Figure 8. Method of recovering the file**

2) Extraction of texts and images

Another method of extracting residual information is to directly extract text and images. Since the rules of storing text and images in PDF files are presented in [3], the contents can be extracted directly.

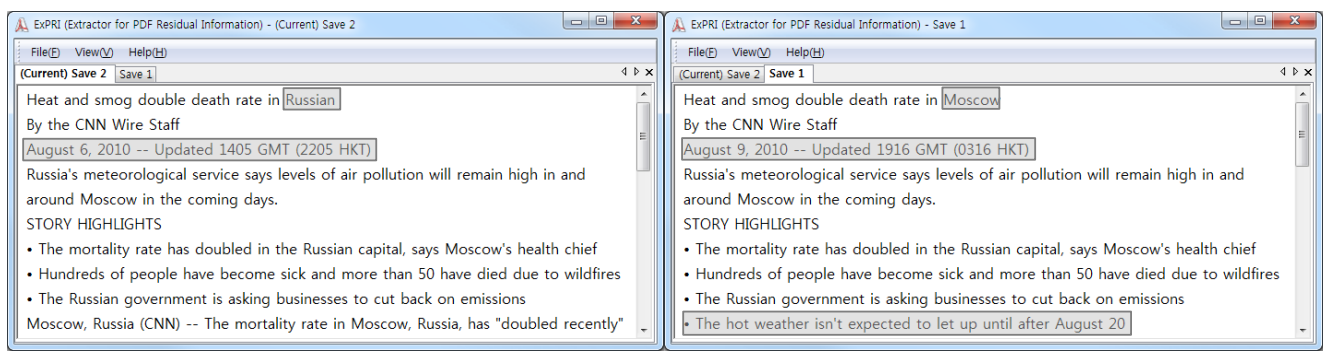

**Figure 9. ExPRI (Extractor for PDF Residual Information)**

In case of the text data, there are two types of ASCII and Unicode. The text data can be extracted by decompressing (or decoding) the data of a target object. Although decompressed (or decoded) data is acquired, it is difficult to extract exact text data. Since the PDF file saves the text with some information such as position, shape and so on, it is necessary to do additional parsing process.

Also, the images in the content can be directly extracted because images are stored as their own image formats.

3) ExPRI (Extractor for PDF Residual Information)

To investigate the forensic attributes of Adobe PDF file, ExPRI(Extractor for PDF Residual Information) has been developed. It is difficult for investigator to change manually byte offset using methodology in Figure 8.

The ExPRI is able to recovery previous files and to extract the text at each storing point. Left side in Figure 9 shows result of extracting modified PDF file's contents. Right side in Figure 9 shows the result of extracting recovered PDF file's contents. Modified and recovered file in Figure 7 demonstrate that ExPRI extracts text correctly.

6. Data hiding method using file-update mechanism

1) Data hiding technique 1

The actually viewed section is the part of 'valid data' of the modified file presented in Figure 10. It means that although the area of 'hidden data' exists inside the file, it is not used. This area can be used to hide data. Even though 'hidden data' is stored in a PDF file, the data cannot be identified using a PDF file viewer. Since the data is compressed (or encoded), it is not possible to verify such data using a simple method like Strings. Therefore, it can be used as a method that simply hides huge amounts of data.

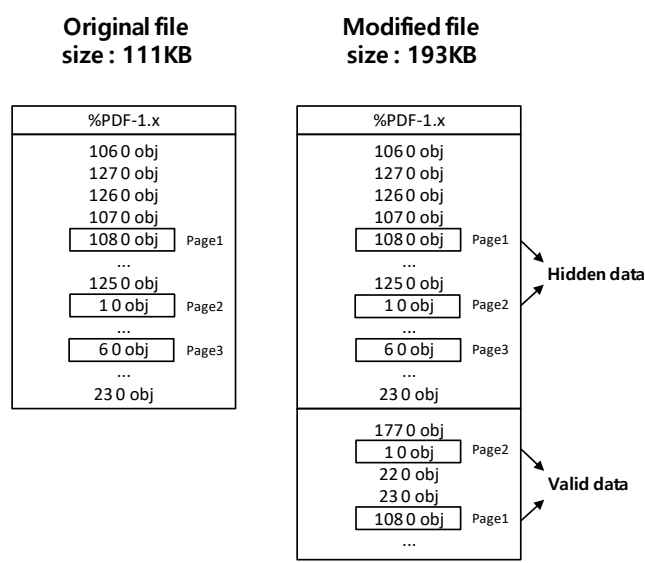

**Figure 10. Hidden and valid data in the modified file**

2) Data hiding technique 2

Because the area of 'hidden data' presented in Figure 10 is not used, the area can be used for hiding the data. Figure 11 illustrates the example of such a way. As the data is hidden using this method, it is not possible to verify the data by extracting the residual information. Also, it is very strong way to hide data with encryption algorithms. Thus, it is very difficult to investigate such hidden data.

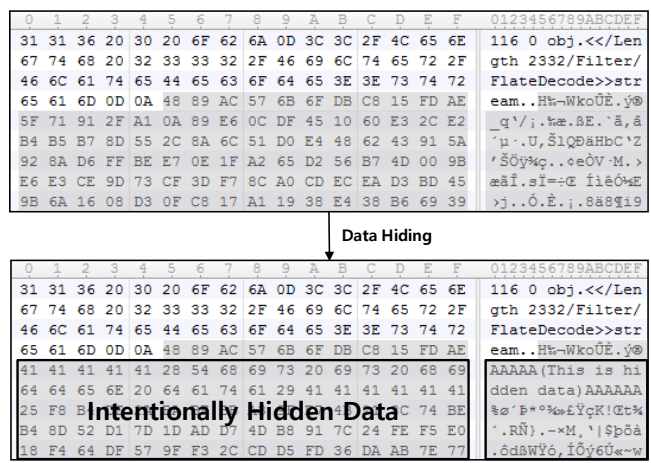

**Figure 11. Data hiding technique 2**

These two methods presented above represent an advantage that is able to hide huge amounts of data compared with the method proposed by Shangping Zhong et al. [4].

7. Conclusion

In digital forensic investigation, there are some cases which documents files are analyzed as evidences. Because researches on electronic document forensics are not enough until now, the analyses have been focused on the contents that can be easily verified by using specific applications. However, this method is insufficient for analyzing PDF files because there are 'hidden data' inside them. In this vein, it is expected that the method introduced in this study is useful to investigate Adobe PDF files. Also, this study will help to improve the admissibility of electronic documentary evidence.

In this study, we analyzed the structure and attributes of Adobe PDF files in a digital forensic viewpoint. It is possible to trace the previous work in a file using its residual information remained in the file based on the update mechanism of an Adobe PDF editing program. In addition, because the area of residual information is not identified by a PDF file viewer, it can be used as a way of data hiding.

For future study, the tool that has been developed at the present time will be completed. It can be used as a forensic analysis module for PDF files because it includes previous file recovery, text and image extraction, and metadata extraction functions for each storing point. In addition, a study on the method that intentionally hides data to the unused area where residual information are stored will be deeply conducted.

References

[1] Matthew Ryan Davis, "Faith in the Format: Unintentional Data Hiding in PDFs," 757Labs.com, 2008.
[2] Jungheum Park, Sangjin Lee, "Forensic investigation of Microsoft PowerPoint files," Digital Investigation, Vol 6, p16~24, 2009.
[3] Adobe Systems, "Document management - Portable document format Part 1: PDF 1.7, First Edition," 2008.
[4] Shangping Zhong, Xueqi Cheng, Tierui Chen, " Data Hiding in a Kind of PDF Texts for Secret Communication, International Journal of Network Security", Vol 4, No 1, P17-26, 2007.
[5] Didier Stevens, Solving a Little PDF Puzzle, URL:http://blog.didierstevens.com/2008/05/07/solving-a-little-pdf-puzzle/, 2008.